\documentclass[12pt]{article}

\usepackage{blkarray}
\usepackage{multirow}
\usepackage{graphicx}
\usepackage{latexsym}
\usepackage{amsmath}
\usepackage{float}

\newcommand{\wt}{\widetilde}
\newcommand{\beq}{\begin{eqnarray}}
\newcommand{\eeq}{\end{eqnarray}}

\numberwithin{equation}{section}

\begin{document}
\title{ Canonical structure \\ of higher derivative theories}
\author{ Eran Avraham, Ram Brustein \ \\ \ \\
Department of Physics, Ben-Gurion University, \\ Beer-Sheva 84105, Israel \\ \hbox{\normalsize E-mail: eranav@bgu.ac.il, ramyb@bgu.ac.il} }
\date{}
\maketitle


\begin{abstract}
The canonical structure of theories whose Lagrangian contains higher powers of time derivatives is often obscured by the nonlinear relationship between the velocities and momenta.   We use the Dirac formalism and define a generalized Legendre transform to overcome some of the difficulties associated with inverting the relation between velocities and momenta. We are then able to  define a standard single valued symplectic structure on phase space and a compatible single valued Hamiltonian. We demonstrate the application of our formalism in several examples.

\end{abstract}

\pagebreak
\section{INTRODUCTION}
Theories containing higher derivatives (HD) appear in physics in
several contexts. The simple cases when the Lagrangian is quadratic in velocities are only approximate, valid at small velocities. These simple Lagrangians  can be viewed as the leading order expansion of more general effective Lagrangians. In field theories, in addition to higher time derivatives, one also encounters higher space derivatives.  In relativistic field theories, the fields are a function of spacetime and the expansion is in spacetime derivatives.

Not all theories containing higher derivatives are sensible. If their equations of motion contain terms with more than two time derivatives, then they posses runaway solutions, signaling an instability at the shortest time scales. We are not interested in such theories. So, we will limit ourselves to theories whose equations of motion do not exhibit such instabilities. These theories have a Lagrangian that is typically a polynomial in velocities and does not depend explicitly on time derivatives of the velocities.

The canonical formulation of HD theories is an important step toward consistently quantizing them. The Ostrogradsky method \cite{ost1} (See also \cite{ost2}) enables one to canonically formulate  a wide class of HD theories and to derive a Hamiltonian for them.  We discuss a class of theories to which the Ostrogradsky method cannot be applied.  As mentioned above, our focus in the paper is on theories containing high powers of the velocities. In this case there are too many solutions to the inversion equation connecting velocities and momenta, leading to a multivalued Hamiltonian. This situation cannot be handled in the Ostrogradsky method because (i) The Lagrangian does not contain new degrees of freedom and (ii) The inversion of the velocities in terms of momenta is multivalued.

Over the years, it became clear that when attempting this task one encounters various difficulties. The issue that we would like to understand is whether the difficulties are fundamental and result from some basic obstruction to applying the canonical formalism to such theories. If this conclusion is verified, it might indicate that the canonical formalism itself is limited in some ways. Alternatively, the difficulty could be technical, specific to some subclasses of such theories and  originate from their added complexity. The latter possibility indicates that we can apply the canonical formalism in general and develop approximation schemes when its application becomes technically difficult. Our results  support this case.

We address, in particular, the problem of finding a Hamiltonian for the class of HD theories of interest.  The method that we have found to be suitable for this task is the Dirac formalism. In this formalism, one extends phase space by adding new variables and then imposes constraints to remove the additional variables. This method allows us to define  good phase space coordinates in which one can solve the constraints and reduce the number of variables  to the original number of degrees of freedom. The end result is  a single valued Hamiltonian which is compatible with the symplectic structure on phase space.

We find that the obstructions are technical and not fundamental. They are equivalent to the difficulty of solving nonlinear equations in standard theories with complicated interactions that involve only the coordinates.

One of the main motivations for the current investigations is to understand the canonical structure of theories of gravity. There, the expansion of the effective Lagrangian is in terms of curvature invariants.
In this context we wish to understand the canonical structure of a class of HD theories of gravity, such as $f(R)$ gravity and Lovelock gravity [\ref{Lovelock}]. Such theories pose difficulties for implementing the Hamiltonian formalism. The high powers of velocities in the Lagrangian lead to a complicated algebraic equation connecting the  canonical momentum and the velocity $P=\frac{\partial L}{\partial \dot{x}}$. This in turn leads to a multivalued Hamiltonian in terms of $P$. This issue was addressed several times in the literature [\ref{MTJ},\ref{CTC},\ref{BQ},\ref{ZYX},\ref{HH},\ref{BHS},\ref{GL}] using different approaches.

The plan of the paper is as follows. In Sec.~\ref{sec.II} we present our formalism for the case of a single variable. We explain how to use the Dirac formalism and how to define the generalized Legendre transform.  We  then discuss several examples.  In Sec.~\ref{sec.III} we extend the formalism for the case of several variables. We conclude by presenting a summary of results and conclusions.

\section{GENERALIZED LEGENDRE TRANSFORM
 FOR A SINGLE VARIABLE}
\label{sec.II}
In this section we describe how to use the Dirac formalism for HD theories and how to choose an appropriate set of coordinates in phase space.

\subsection{Formalism}

Let us consider the Lagrangian $L(x,\dot{x})$ of a single dynamical variable. Rather than treating $\dot{x}$ as the time derivative of $x(t)$,  we wish to treat $\dot{x}$ as an independent variable. The new independent variable will be labeled by $Q$. To this end we add a Lagrange multiplier $\lambda$ that imposes the equality of $Q$ and $\dot{x}$ at all times. Then the new Lagrangian is given by
\beq
\wt{L}=L(x,Q)+\lambda(\dot{x}-Q).
\eeq
Because the resulting Lagrangian is singular: the momentum of $Q$ and $\lambda$ vanishes,  we  turn to the Dirac formalism [\ref{dirac}]. Performing a Legendre transform we obtain the canonical Hamiltonian,
\beq
H_C=\lambda Q-L(x,Q)
\eeq
and three primary constraints,
\beq
\phi_1 = P_Q, \quad
\phi_3 = P_\lambda, \quad
\phi_4 = P_x-\lambda .
\eeq
The constraints $\phi_{3}$, $\phi_{4}$  are not dynamical and can be harmlessly substituted into the modified Hamiltonian (see [\ref{david}]),
\beq
{H}_1&=& H_C+u_1\phi_1 \cr
&=& P_xQ-L(x,Q)+u_1P_Q.
\eeq
Demanding consistency from the constraint $\{\phi_1,{H}_1\}=-\partial_Q {H}_1$ we find a secondary one
\beq
\phi_2=P_x-\partial_QL.
\eeq
Including this constraint in the Hamiltonian results in the total Hamiltonian,
\beq
H_T=P_xQ-L(x,Q)+u_1P_Q+u_2(P_x-\partial_QL).
\eeq

On shell, we may set the constraints to zero, provided that we use the Dirac  brackets rather than the Poisson brackets [\ref{dirac}].
The Poisson brackets between the two second class constraints  are given by
\beq
\{\phi_1,\phi_2\}=-\partial_Q\phi_2=\partial^2_QL(x,Q),
\eeq
so the Dirac matrix and its inverse are given by
\beq
M_{ij}=\partial^2_QL(x,Q)\begin{pmatrix}
 0 & 1 \\
 -1 & 0  \\
 \end{pmatrix},\quad
 M^{-1}_{ij}=-\frac{1}{\partial^2_QL(x,Q)}\begin{pmatrix}
 0 & 1 \\
 -1 & 0  \\
 \end{pmatrix}.
\eeq
Then the Dirac brackets for any pair of physical quantities $A$, $B$, is given by
\beq\nonumber
\{A,B\}_D=\{A,B\}+ \frac{1}{\partial^2_QL(x,Q)} \Big[\{A,\phi_1\}\{\phi_2,B\}-\{A,\phi_2\}\{\phi_1,B\}\Big].
\eeq

We can now calculate  the final Hamiltonian
\beq
H_F=Q\partial_Q L(x,Q)-L(x,Q).
\eeq
However, this Hamiltonian is valid with the caveat that the Dirac bracket between phase space variables is not given by the standard expression  $\{x,Q\}_D=1$, rather it is given by the inverse of the Hessian,
\beq
\quad \{x,Q\}_D=\left({\partial^2_Q L(x,Q)}\right)^{-1}.
\eeq
To proceed we need to change coordinates in phase space: we need to find some new functions, say $f(x,Q)$ and $g(x,Q)$ such that
$\{f,g\}_D=1$.
  Noting that $\phi_1$ acts like $\partial_Q$ and $\phi_2$ like $\partial_x$, we can write the condition explicitly:
\beq
\partial_xf\partial_Qg-\partial_Qf\partial_xg=\partial^2_QL(x,Q).
\label{genpoisson}
\eeq
Once two such functions are found, we need to invert the relations between $(x,Q)$ and $(f,g)$ and reexpress the final Hamiltonian in terms of new canonical variables.

When a pair of functions $(f,g)$ that satisfies condition (\ref{genpoisson}) is found, it is possible to use the standard Poisson brackets with respect to new basis $(f,g)$ rather than the Dirac brackets. This follows from the chain rule property that both the Poisson and Dirac brackets posses:
\beq
\dot{f}(x,Q)&=&\biggl\{f(x,Q),H\left.\bigl(f(x,Q),g(x,Q)\right.\bigr)\biggr\}_D \cr \vbox{\vspace{.3in}}
&=&\{f,f\}_D\frac{\partial H}{\partial f}+\{f,g\}_D\frac{\partial H}{\partial g} =
\frac{\partial H}{\partial g},
\eeq
where the last equality is due to condition (\ref{genpoisson}) which guarantees that $\{f,g\}_D=1$. The final result is  therefore
\beq
\dot{f}(x,Q)=\{f,H\}_{(f,g)}.
\eeq

The standard definition of canonical conjugates $f=x$,  $g=\frac{\partial L}{\partial \dot{x}}=\frac{\partial L}{\partial Q}$, is
a particular case of a more general relation (\ref{genpoisson}). This freedom to choose a different set coordinates in phase space can be used to bypass some of the difficulties one may encounter using the standard Legendre transform.

\subsection{A simple example}

Consider the simple HD Lagrangian
\beq
L=\frac{1}{4}\dot{x}^4-\frac{1}{2}k\dot{x}^2-\frac{1}{2}\omega x^2,
\label{Lx^2}
\eeq
where $k,\omega>0$ are constants. A similar Lagrangian (with $\omega=0$) was discussed recently in [\ref{MTJ},\ref{CTC},\ref{BQ},\ref{ZYX},\ref{HH}].

If we perform the standard Legendre transform we encounter the cubic equation for the momentum $P_x$ in terms of the velocity $\dot{x}$,
\beq
P_x= \dot{x}^3-k\dot{x}.
\eeq
The inversion of $\dot{x}$ in terms of $P_x$ leads to a  multivalued Hamiltonian, with cusps at minima (see Fig.~1). The lowest energy solution of system ``spontaneously breaks time translation" [2], because at the cusps the velocity  is nonvanishing $\dot{x}=\pm\sqrt{k/3}$.
\begin{figure}[H]
\begin{center}
\includegraphics[scale=0.5]{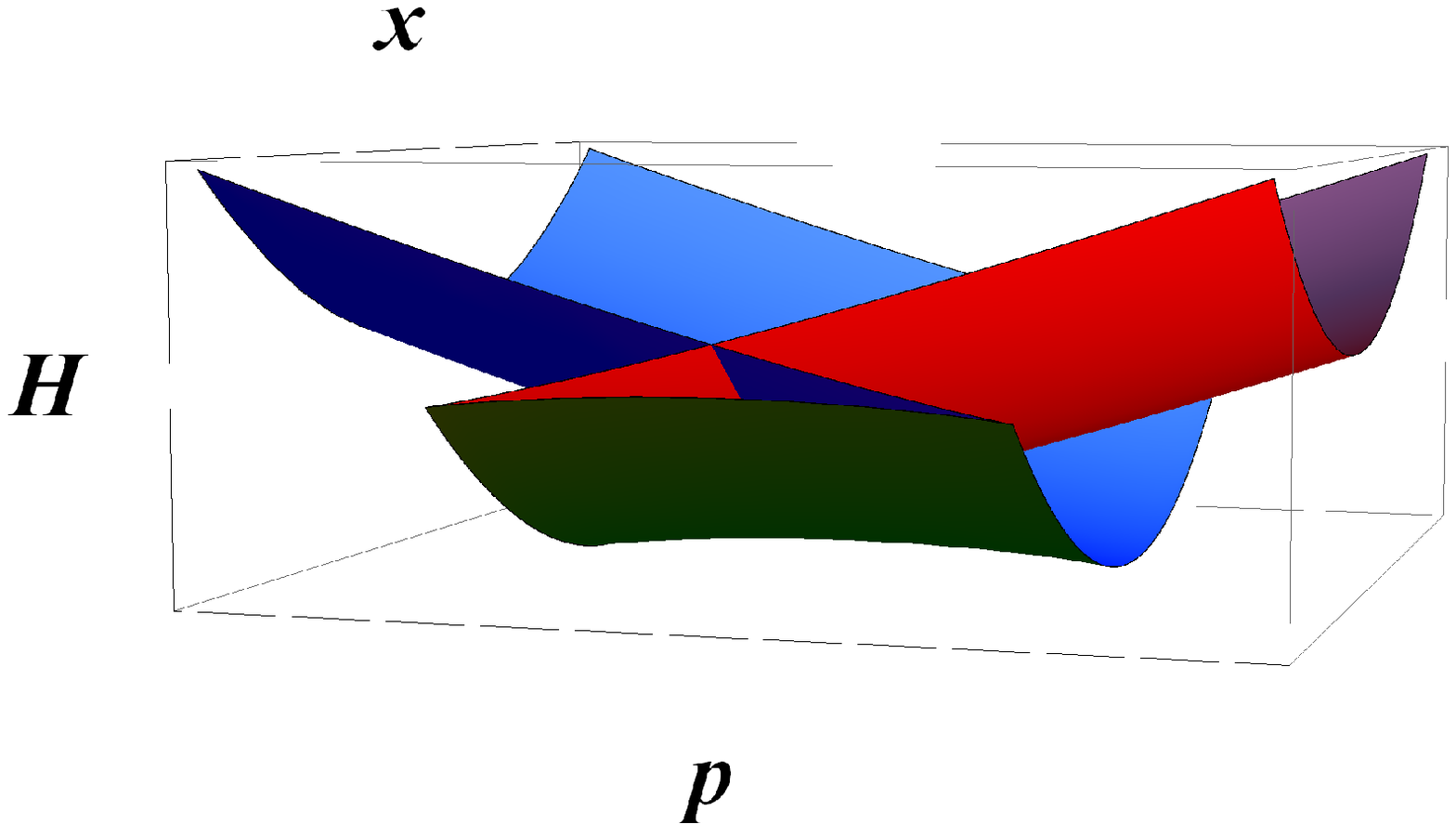}\\
Figure 1: Multivalued Hamiltonian
\end{center}
\end{figure}
Applying the generalized Legendre transform procedure that we have described in the previous subsection proceeds as follows: We change variables to $(x,Q)$
\beq
L\rightarrow \wt{L}= \frac{1}{4} Q^4-\frac{1}{2}k Q^2-\frac{1}{2}\omega x^2+\lambda(\dot{x}-Q).
\eeq
The corresponding canonical Hamiltonian is
\beq
H_C=\frac{3}{4} Q^4-\frac{1}{2}k Q^2+\frac{1}{2}\omega x^2
\eeq
and the two second class constraints are
\beq
 \phi_1=P_Q, \quad
 \phi_2=P_x -Q^3+kQ.
\eeq
We turn to find $f$ and $g$, satisfying the condition
\beq
\partial_x f\partial_Q g-\partial_Q f\partial_x g&=&\partial_Q^2 L(x,Q) \cr
&=& 3 Q^2-k\,
\eeq
Choosing $g = Q$ determines $f$ up to some function  $\psi(Q)$ which we set to zero
\beq
f=x(3 Q^2-k).
\label{1}
\eeq
Now the inversion is simple
\beq
Q &=& g,
\label{OV1}\\
x &=& \frac{f}{3 Q^2-k}=\frac{f}{3 g^2-k}.
\label{OV2}
\eeq
Substituting $x$ and $Q$ in terms of $f$ and $g$, we find the  final  Hamiltonian which is single valued and now has two standard canonical conjugate variables satisfying the standard Poisson brackets, $\{f,g\}=1$,
\beq
H(f,g)=\frac{\omega}{2(3g^2-k)^2} f^2+\frac{3}{4} g^4-\frac{1}{2}k g^2.
\label{hfg}
\eeq
The resulting Hamiltonian effectively describes a particle in a quartic potential with a divergent mass term.

The Hamilton equations are given by
\beq
\dot{g}&=&-\frac{\partial H}{\partial f}=-\frac{\omega}{(3g^2-k)^2} f,\\ \dot{f}&=&\frac{\partial H}{\partial g}= 3g^3-kg  -6\omega \frac{g f^2}{(3 g^2 - k)^3} .
\eeq
Which, using Eqs.~(\ref{OV1}--\ref{OV2}), reproduce the same equations of motion derived from the starting Lagrangian (\ref{Lx^2}),
\beq
(3\dot{x}^2-k)\ddot{x}=-\omega x.
\eeq
The Hamiltonian of Eq.~(\ref{hfg}) is plotted in Fig.~2. It has two minima at $f=0,\quad g=\pm\sqrt{k/3}$. So the expected nonvanishing values of the velocities are now a result of minimizing the Hamiltonian rather than the branched nature of phase space.

\begin{figure}[H]
\begin{center}
\includegraphics[scale=0.6]{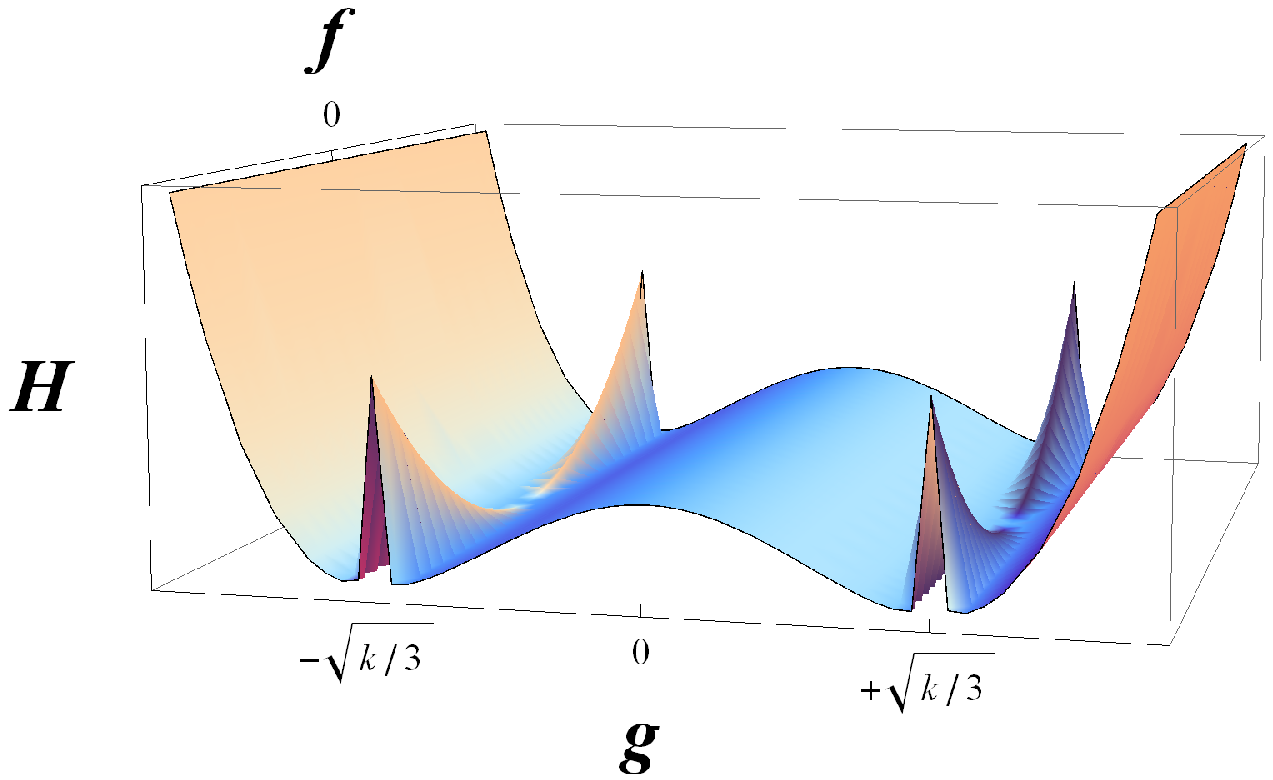}\\
Figure 2: Single valued Hamiltonian
\end{center}
\end{figure}

\subsection{Another example: Minisuperspace Gauss-Bonnet}
As explained in the Introduction, HD theories appear in the context of generalized theories of gravity. There, higher curvature terms added to the action corresponds to high powers of ``velocity." In the following  we will derived the Hamiltonian for a simple Gauss-Bonnet cosmological model.
We do not include here the lapse function and the associated Hamiltonian constraint. This is discussed, for example, in [\ref{mena1}],[\ref{mena2}].

The Gauss-Bonnet Lagrangian, is given by
\beq L_{GB}=\sqrt{-g}[R+\gamma (R^2-4R_{\alpha\beta}^2+R_{\alpha\beta\gamma\delta}^2)]
\label{LGB1}
\eeq
Assuming a metric of the form
\beq
ds^2=-dt^2+a(t)^2dx_i^2
\label{metric}
\eeq
in $D=5$ with zero spatial curvature, we use (\ref{metric}) in (\ref{LGB1})
 and integrate by parts (see [\ref{mena1}],[\ref{mena2}]) to find the Gauss-Bonnet minisuperspace Lagrangian,
\beq
L_{GB}=\frac{1}{2}a^2\dot{a}^2+\frac{1}{4}\alpha\dot{a}^4.
\label{LGB2}
\eeq
As in the previous example, standard Legendre transform will lead to a multivalued Hamiltonian.
Applying our method, we first need to change variables $a\rightarrow e^{x}$, changing (\ref{LGB2}) to
\beq
\wt{L}_{GB}=e^{4x}[\frac{1}{2}\dot{x}^2+\frac{1}{4}\alpha\dot{x}^4].
\eeq
Adding the Lagrange multiplier and switching to the $(x,Q)$ coordinates we find the corresponding canonical Hamiltonian,
\beq
H_C=e^{4x}[\frac{1}{2}Q^2+\frac{3}{4}\alpha Q^4].
\label{HCGB}
\eeq
Solving the generalized Legendre equation (\ref{genpoisson}) requires that
\beq
\partial_x f\partial_Q g-\partial_Q f\partial_x g = e^{4x}[1+3\alpha Q^2].
\eeq
We choose $g=Q$, and find that
\beq
f=\frac{1}{4}e^{4x}[1+3\alpha Q^2].
\eeq
Inverting $(x,Q)$ in terms of $(f,g)$ and substituting into Eq.~(\ref{HCGB}) we obtain the final Hamiltonian
\beq
H(f,g)=fg^2\Big[1+\frac{1}{3\alpha g^2+1}\Big].
\eeq

If $\alpha<0$  the minimum of the Hamiltonian lies in $g=\pm\sqrt{\frac{2}{3\mid \alpha \mid}}$ corresponding to $a=e^{\pm\sqrt{\frac{2}{3\mid \alpha \mid}}t}$, that is, an expanding or contracting universe.

\section{THE GENERALIZED LEGENDRE TRANSFORM FOR SEVERAL VARIABLES}
\label{sec.III}
Next we generalize our formalism for the case of several variables.
\subsection{Formalism}

Our starting point is now a Lagrangian which is a functional of several variables $L(x_i,\dot{x}_i)$. As in the previous section, we add the Lagrange multipliers
\beq
L(x_i,\dot{x}_i) \rightarrow L(x_i,Q_i) +\sum_i \lambda_i(\dot{x}_i-Q_i).
\eeq
The corresponding canonical Hamiltonian is given by
\beq
H_C=\sum_i P_{x_i}Q_i-L(x_i,Q_i).
\eeq
The primary and secondary constraints are the following,
\beq
\phi^1_i = P_{Q_i}, \quad
\phi^2_i = P_{x_i}-\frac{\partial L}{\partial Q_i}
\eeq
and their Poisson brackets are the following,
\beq
&&\{\phi^1_i,\phi^1_j\}=0,\\
&&\{\phi^1_i,\phi^2_j\}=\frac{\partial^2 L}{\partial Q_i \partial Q_j}\equiv W_{ij},\label{W}\\
&& \{\phi^2_i,\phi^2_j\}=\frac{\partial^2 L}{\partial x_i \partial Q_j}-\frac{\partial^2 L}{\partial x_j \partial Q_i}\equiv B_{ij}\label{B}.
\eeq
The Dirac matrix and its inverse can be expressed in terms of the matrices $W_{ij}$ and $B_{ij}$,
\begin{eqnarray}\def\arraystretch{0.5}
M_{IJ}=
     \begin{blockarray}{(ccc|ccc)l}
        &&&\BAmulticolumn{3}{c} {\multirow{3}{*}{$W_{ij}^{-1}$}}&\multirow{3}{*}{}\\&0&\\&&&\\
        \cline{1-6}
        \BAmulticolumn{3}{c|} {\multirow{3}{*}{$-W_{ij}^{-1}$}}&\multirow{3}{*}{}\\&&&&{ B_{ij}}&&\\&&&&&&\\
    \end{blockarray}, \quad
  M_{IJ}^{-1}=
    \begin{blockarray}{(ccc|ccc)l}
        &&&\BAmulticolumn{3}{c} {\multirow{3}{*}{$-W_{ij}^{-1}$}}& \multirow{3}{*}{}\\&W_{ik}^{-1}B_{kl}W_{lj}^{-1}&\\&&&\\
        \cline{1-6}
        \BAmulticolumn{3}{c|} {\multirow{3}{*}{$W_{ij}^{-1}$}}&\multirow{3}{*}{}\\&&&&{ \huge {0}}&&\\&&&&&&
    \end{blockarray}.
\end{eqnarray}
Thus the Dirac Brackets are given by
\beq
&&\{f(x,Q),g(x,Q)\}_D= \{f,g\}-\sum_{IJ}\{f,\phi_I\}M_{IJ}^{-1}\{\phi_J,g\} \cr
&=&
\def\arraystretch{0.5}
0-\sum_{ij}
\begin{pmatrix}
\dots,\frac{\partial f}{\partial Q_i},\dots,\frac{\partial f}{\partial x_i},\dots
\end{pmatrix}
\begin{matrix}
 \begin{blockarray}{(ccc|ccc)l}
        &&&\BAmulticolumn{3}{c} {\multirow{3}{*}{$-W_{ij}^{-1}$}}& \multirow{3}{*}{}\\&W_{ik}^{-1}B_{kl}W_{lj}^{-1}&\\&&&\\
        \cline{1-6}
        \BAmulticolumn{3}{c|} {\multirow{3}{*}{$W_{ij}^{-1}$}}&\multirow{3}{*}{}\\&&&&{ \huge {0}}&&\\&&&&&&\\
    \end{blockarray}
    \end{matrix}
    \begin{pmatrix}
\vdots\\
 -\frac{\partial g}{\partial Q_j}    \\
 \vdots\\
   -\frac{\partial g}{\partial x_j} \\
  \vdots\\
 \end{pmatrix}\cr
&=&\sum_{ij}\Bigg[W_{ij}^{-1}\Big(\frac{\partial f}{ \partial x_i}\frac{\partial g}{ \partial Q_j}-\frac{\partial f}{ \partial Q_i}\frac{\partial g}{ \partial x_j}\Big)+W_{ik}^{-1}B_{kl}W_{lj}^{-1}\frac{\partial f}{ \partial Q_i}\frac{\partial g}{ \partial Q_j}\Bigg].
\eeq
We can now write the generalized condition for the canonical pairs
\beq
\{f_a,g_b\}_D &=& \sum_{ij}\Biggl[W_{ij}^{-1}A_{ij}+T_{ij}C_{ij}\Biggr] \cr &=& \text{Tr}(W^{-1}A^T+T C^T)\cr &=& \delta_{ab}.
\label{gc}
\eeq
In the derivation of Eq.(\ref{gc}) we have used  the identity $\sum_{ij} E_{ij}F_{ij} = \text{Tr}(EF^T)$ and introduced the following notations
\beq \nonumber
A_{ij}\equiv \Big(\frac{\partial f_a}{ \partial x_i}\frac{\partial g_b}{ \partial Q_j}-\frac{\partial f_a}{ \partial Q_i}\frac{\partial g_b}{ \partial x_j}\Big),\quad
T_{ij} \equiv W_{ik}^{-1}B_{kl}W_{lj}^{-1}, \quad
C_{ij} \equiv \frac{\partial f_a}{ \partial Q_i}\frac{\partial g_b}{ \partial Q_j}.\\
\eeq
Let us verify that our formalism can accommodate the standard choice of $f_a=x_a$ and $g_a=\frac{\partial L}{ \partial Q_a}$. In this case $A^T$ and $C$ are given by
\beq \nonumber \def\arraystretch{0.5}
 A_{ij}^T=\begin{pmatrix}
0&\dots & \frac{\partial L}{ \partial Q_a \partial Q_1}&\dots & 0\\
\vdots &&\vdots &  &\vdots \\
0&\dots & \frac{\partial L}{ \partial Q_a \partial Q_i}&\dots & 0\\
  \vdots & & \vdots&  &\vdots \\
 \end{pmatrix}=
 \begin{pmatrix}
0&\dots & W_{1a}&\dots & 0\\
\vdots & &\vdots & &\vdots  \\
0&\dots & W_{ia}&\dots & 0\\
 \vdots  & &\vdots & &\vdots  \\
 \end{pmatrix},\quad C_{ij}=0.\\
\eeq
So, we find
\beq\def\arraystretch{0.5}
\{f_a,g_a\}_D=\text{Tr}(W_{ik}^{-1}A_{kj}^T)=\text{Tr} \begin{pmatrix}
0 &\dots & 0&\dots & 0\\
 &\ddots & &   \\
0&\dots & 1&\dots &0\\
  && & \ddots& &   \\
  0 &\dots & 0&\dots & 0
 \end{pmatrix}=1.
\eeq
One can also check that $\{x_a,x_b\}_D,\{x_a,\frac{\partial L}{ \partial Q_b}\}_D,\{\frac{\partial L}{ \partial Q_a},\frac{\partial L}{ \partial Q_b}\}_D=0 $ for $a\neq b$ as expected.

\subsection {The case of $B_{ij}=0$}

If the coupling between the velocities and the coordinates vanishes, then the matrix $B_{ij}=\frac{\partial^2 L}{\partial x_i \partial Q_j}-\frac{\partial^2 L}{\partial x_j \partial Q_i}$ vanishes and condition (\ref{gc}) reduces to
\beq
\{f_a,g_b\}_D=\text{Tr}(W^{-1}A^T)=\delta_{ab}.
\label{rc}
\eeq
In this case, the added complexity is due only to the higher powers of the velocity (say a term of the form $\dot{x}^4$). It is useful to choose $g_a=Q_a$ which determines the form of $A_{kj}^T$,
\beq\def\arraystretch{0.5}
A_{kj}^T=\begin{pmatrix}
0&\dots &\frac{\partial f_a}{  x_1}&\dots & 0\\
\vdots &&\vdots &  &\vdots \\
0&\dots & \frac{\partial f_a}{ \partial x_i}&\dots & 0\\
  \vdots & & \vdots&  &\vdots \\
 \end{pmatrix}.
\eeq
Substituting  into Eq.~(\ref{rc}) we find the condition for $f_a$ is the following
\beq
\frac{\partial f_a}{ \partial x_i} =W_{ia},
\label{cB=0}
\eeq
which has a solution only if $W_{ia}$ is an exact differential.

In order to complete the analysis we prove that all other brackets vanish, i.e
\beq
\{Q_a,Q_b\}_D,\ \{g_a,g_b\}_D,\ \{Q_a,g_b\}_D,\ \{Q_b,g_a\}_D=0.
\eeq
The first condition comes from the definition of $A_{ij}$.
The second from the requirement that $B_{ij}=0$.
For the third and fourth conditions let us write explicitly
\beq\def\arraystretch{0.5}
A_{kj}^T(f_a,Q_b)=
\begin{pmatrix}
0&\dots &\frac{\partial f_a}{x_1}&\dots & 0\\
\vdots &&\vdots &  &\vdots \\
0&\dots & \frac{\partial f_a}{ \partial x_i}&\dots & 0\\
  \vdots & & \vdots&  &\vdots \\
 \end{pmatrix}.
\eeq
Now, the $``b"$ column  is occupied by $ \frac{\partial f_a}{ \partial x_i}$ which equals by construction to $W_{ia}$. This switch of the $``a"$ and $``b"$ columns then implies that $\text{Tr}(W_{ik}^{-1}A_{kj}^T)=0$.

\subsection{An example with $B_{ij}=0$ }
We conclude with an example for which $ B_{ij}=0$.
Consider the following Lagrangian,
\beq
 L=\frac{1}{4}(\dot{x}^2+\dot{y}^2-k)^2-V(x,y).
\eeq
Applying our method from sec. (3.2) we find the Hessian matrix is
\beq
W_{ik}=\begin{pmatrix}
3Q_x ^2+Q_y^2-k & 2Q_xQ_y \\
2Q_xQ_y & 3Q_y ^2+Q_x^2-k
 \end{pmatrix}.
\eeq
fixing $g_1=Q_x$ we find that Eqs.~(\ref{cB=0}) lead to the following equations for  $f_1$,
\beq
\frac{\partial f_1}{\partial x}&=& 3Q_x ^2+Q_y^2-k,\\
\frac{\partial f_1}{\partial y}&=& 2Q_xQ_y.
\eeq
The solution of the previous equations is given by
\beq
f_1 =x(3Q_x ^2+Q_y^2-k)+ 2yQ_xQ_y.
\eeq
Similarly we choose $g_2=Q_y$ and solve for $f_2$
\beq
f_2=y(3Q_x ^2+Q_y^2-k)+ 2xQ_xQ_y.
\eeq
The inversion of $\{x,Q_x,y,Q_y\}$ in terms of $\{f_1,g_1,f_2,g_2\}$ is  given  by
\beq
x &=& \frac{-2 f_2 g_1 g_2 + f_1 (3 g_1^2 + g_2^2 - k)}{
9 g_1^4 + 2 g_1^2 (g_2^2 - 3 k) + (g_2^2 - k)^2},\\
Q_x&=&g_1,\\
y &=& \frac{-2 f_1 g_1 g_2 + f_2 (3 g_1^2 + g_2^2 - k)}{
9 g_1^4 + 2 g_1^2 (g_2^2 - 3 k) + (g_2^2 - k)^2},\\
Q_y&=&g_2,
\eeq
and the last step is to substitute back into the canonical Hamiltonian and express it in terms of the new variables $f_i$ and $g_i$,
\beq
 H(f,g)= \frac{3}{4}(g_1^2+g_2^2-\frac{1}{3}k)^2 +V[x(f_i,g_i),y(f_i,g_i)]
\eeq
Similar to the first example, the potential $V(x,y)$ plays the role of a complicated kinetic term, while the original kinetic term becomes a Mexican hat potential.

\section{CONCLUSIONS}

Theories whose Lagrangian contains high powers of the velocities seem to lead to a multivalued canonical description. However, we have seen that this conclusion is not necessary valid in many cases. Using the Dirac formalism we were able to derive a generalized definition of the canonical variables. This new definition allows  additional freedom for solving the velocities in terms of the momenta. Our conclusion is that in some cases the multivalued nature of phase space originates in an inappropriate  choice of coordinates rather than from a fundamental limitation.  In some cases, a good choice of phase space coordinates enables one to define a single valued Hamiltonian with a standard, single valued, symplectic structure.

When a standard canonical structure can be found, we expect that the quantization can follow the standard rules, turning coordinates into operators and solving the corresponding Schr\"{o}dinger equation. If one so wishes, one can transform the result back to the original coordinates by applying the formalism of canonical transformation in quantum mechanics [\ref{canonical}].
However, one will have to address quantum ordering issues
when defining  the quantum Hamiltonian. Either the
Poisson or Dirac brackets are noncanonical, or the Hamiltonian possesses
quantum ordering ambiguities. We hope to discuss the quantization of higher-derivative theories in a future publication.

In the general case, we expect that  finding good phase space coordinates is not always possible. For example, if we attempt to apply our formalism in the case of a more complicated Lagrangian, say,  $L=ax^3\dot{x}^2 +b\dot{x}^3 +cx^5\dot{x}^4 -V(x)$. We will be able to find $f$ and $g$  satisfying Eq.~(\ref{genpoisson}). But for every solution the inversion of $(x,Q)$ in terms of $(f,g)$ will be multivalued. In such cases, for which the Hamiltonian is truly multivalued, one needs a different approach. It is clear, however, that these cases are not fundamentally flawed in any way.

Even though we were able to find a Hamiltonian for the minisuperspace Gauss-Bonnet action, trying to apply the formalism for the general case will face  difficulties. This is because  in the case of a Lagrangian with $N$-degrees of freedom, the number of equations one needs to solve  scales as $\binom{2N}{2}$. This technical difficulty is greatly reduced if the coupling between the coordinates and the velocities vanishes. In the general Gauss-Bonnet action, the coordinates: the metric and the velocities: the extrinsic curvature,  are coupled in a nontrivial way. Progress in this direction will require  the construction of an efficient method to handle the growing number of equations.

\section*{Acknowledgments} We thanks Aharon Davidson for helpful discussions and Jose Edelstein for comments on the manuscript. The research  was supported by the Israel Science Foundation Grant No. 239/10.

\end{document}